\documentclass[aps,prb,twocolumn,superscriptaddress,showpacs]{revtex4}
\usepackage{graphicx}% Include figure files
\usepackage{dcolumn}% Align table columns on decimal point
\usepackage{bm}% bold math
\usepackage{amsmath}

\usepackage{color}

\begin{document}
%\preprint{}

\title{Atomic displacements and lattice distortion in the magnetic-field-induced charge ordered state of SmRu$_{4}$P$_{12}$
}


\author{Takeshi Matsumura}
\email[]{tmatsu@hiroshima-u.ac.jp}
%\thanks{}
\affiliation{Department of Quantum Matter, AdSM, Hiroshima University, Higashi-Hiroshima 739-8530, Japan}
\affiliation{Institute for Advanced Materials Research, Hiroshima University, Higashi-Hiroshima 739-8530, Japan}
\author{Shinji Michimura}
\affiliation{Department of Physics, Faculty of Science, Saitama University, Saitama 338-8570, Japan}
\author{Toshiya Inami}
\affiliation{Synchrotron Radiation Research Center, National Institutes for Quantum and Radiological Science and Technology, Sayo, Hyogo 679-5148, Japan}
\author{Kengo Fushiya}
\affiliation{Department of Physics, Tokyo Metropolitan University, Hachioji, Tokyo 192-0397, Japan}
\author{Tatsuma D. Matsuda}
\affiliation{Department of Physics, Tokyo Metropolitan University, Hachioji, Tokyo 192-0397, Japan}
\author{Ryuji Higashinaka}
\affiliation{Department of Physics, Tokyo Metropolitan University, Hachioji, Tokyo 192-0397, Japan}
\author{Yuji Aoki}
\affiliation{Department of Physics, Tokyo Metropolitan University, Hachioji, Tokyo 192-0397, Japan}
\author{Hitoshi Sugawara}
\affiliation{Department of Physics, Kobe University, Kobe 657-8501, Japan}


\date{\today}

\begin{abstract}
Structural properties of SmRu$_4$P$_{12}$ in the anomalous magnetic ordered phase between $T^*\sim 14 $ K and $T_{\text{N}}=16.5$ K in magnetic fields has been studied by x-ray diffraction. 
Atomic displacements of Ru and P, reflecting the field-induced charge order of the $p$ electrons, have been deduced by analyzing the intensities of the forbidden Bragg peaks, assuming a cubic space group $Pm\bar{3}$. 
Also, by utilizing high-resolution x-ray diffraction experiment, we observed a splitting of fundamental Bragg peaks, clarifying that the unit cell in the magnetic ordered phase is rhombohedral elongated along the $[1\, 1\, 1]$ axis. 
Responses of the rhombohedral domains to the magnetic field, which reflects the direction of the magnetic moment, is studied in detail.  
\end{abstract}

\pacs{
71.27.+a %Strongly correlated electron systems; heavy fermions
, 71.30.+h %Metal-insulator transitions and other electronic transitions
, 75.25.Dk %Orbital, charge, and other orders, including coupling of these orders
, 61.05.cp %X-ray diffraction in crystal structure
}
%\keywords{}

\maketitle


\section{Introduction}
Coexistence of magnetic and orbital degrees of freedom in $f$ electron systems gives rise to a rich variety of ordering phenomena such as multipole orderings and have attracted longstanding interest.\cite{Kuramoto09,Santini09} 
The inter-ionic interaction here is normally mediated by hybridization with the itinerant electrons. 
For $f$ electrons to hybridize with the itinerant electrons, the symmetry relation between the $f$-electron crystal field (CF) state and the itinerant electron state is of fundamental importance. 
Although this fact is often neglected, the CF-state dependent hybridization, in general, should play an important role in the ordering phenomena. 
This is actually the case in filled skutterudite compounds RT$_{4}$X$_{12}$ (R=rare earth, T=transition metal, and X=P, As, and Sb).\cite{Maple07,Shiina12}

A typical case is the metal-insulator transition in PrRu$_4$P$_{12}$, which is accompanied by a charge ordering (CO) of the $p$ electrons and a staggered ordering of the $f$-electron CF states.\cite{Sekine97,Lee04,Iwasa05a} 
The main conduction $p$-band of the RT$_{4}$X$_{12}$ compounds consists of the $xyz$-type ($a_u$-type) molecular orbitals of X$_{12}$ icosahedra, forming a body-centred-cubic lattice of $Im\bar{3}$ space group. 
This $p$ band has a strong nesting instability with $\bm{q}=(1, 0, 0)$, favoring the CO state.\cite{Harima03,Harima08} 
However, this $p$ band alone cannot account for the CO. 
The phase transition is realized by using the electronic degrees of freedom of $f$ electrons with the singlet-triplet CF levels through the CF-state dependent $p$-$f$ hybridization. 
Only the translational symmetry of $Im\bar{3}$ is broken, leaving the local symmetry of Pr unchanged in the low temperature phase of $Pm\bar{3}$. 
Since this CO accompanies a staggered ordering of the CF states, the ordering is also called an antiferro-hexadecapole order, or a scalar order, with the $\Gamma_1$ totally symmetric representation.\cite{Takimoto06,Kiss06,Shiina10} 
A similar ordering of totally symmetric order parameter is also observed in PrFe$_4$P$_{12}$, in which the $d$-$f$ hybridization also needs to be taken into account.\cite{Shiina12a,Shiina12b} 

SmRu$_4$P$_{12}$ has also attracted interest because of its mysterious ordered phase below the antiferromagnetic (AFM) dipole transition at $T_{\text{N}}=16.5$ K, where a metal-insulator transition also takes place.\cite{Sekine00,Matsuhira02,Sekine03,Matsuhira05,Sekine05,Matsunami05,Kikuchi07a} 
The AFM dipole order has been well established by muon spin relaxation,\cite{Hachitani06a,Ito07} nuclear magnetic and quadrupole resonance (NMR/NQR),\cite{Hachitani06b,Masaki06,Masaki07,Masaki08} nuclear resonant forward scattering,\cite{Tsutsui06} and neutron diffraction.\cite{Lee12} 
What is intriguing is that another phase transition develops at $T^*\sim 14$ K by applying a magnetic field in the AFM phase. 
The intermediate phase $(T^{*}<T<T_{\text{N}})$ expands, i.e., $T_{\text{N}}$ increases and $T^{*}$ decreases, with increasing the field.\cite{Matsuhira02,Sekine03,Matsuhira05,Sekine05} 
The paramagnetic phase above $T_{\text{N}}$, the intermediate phase, and the low temperature phase below $T^{*}$ have been named phase I, II, and III, respectively. 
Initially, to interpret this intermediate phase, a possibility of magnetic octupole order was proposed as a primary order parameter.\cite{Yoshizawa05,Aoki07,Kiss09} 
However, in spite of intensive experimental studies, the microscopic nature of the ordered phases in SmRu$_4$P$_{12}$ has remained unresolved. 

A microscopic model to explain this phase was proposed by Shiina.\cite{Shiina13,Shiina14a,Shiina14b} 
The model is based on the $p$-$f$ hybridization between the $\Gamma_8$--$\Gamma_7$ CF levels of $f$ electrons and the $a_u$-type $p$-band of the P$_{12}$ molecular orbitals. 
Since only the $\Gamma_7$ state mixes with the $a_u$ band by symmetry, the AFM interaction occurs only between the $\Gamma_7$ states. Then, although the CF ground state is expected to be the $\Gamma_8$ quartet from the released magnetic entropy of $R\ln 4$ at $T_{\text{N}}$,\cite{Matsuhira05} the AFM ordered state is constructed from the $\Gamma_7$ states. 
In addition to the magnetic exchange interaction, the charge density of $p$ electrons around Sm is coupled with the CF level splitting of Sm. 
When the charge density increases around Sm, the $\Gamma_7$ state prefers to be the ground state. 
Then, since the ordering of the charge density with $\bm{q}=(1,0,0)$ lowers the total energy of the $a_u$ band, a staggered ordering of the CF state also leads to an energy gain. This effect is induced in magnetic fields through the Zeeman splitting of the CF state, resulting in the field-induced CO phase near $T_{\text{N}}$. 

Stimulated by the theory, we have recently performed resonant and nonresonant x-ray diffraction experiments and obtained the results strongly supporting the theory.\cite{Matsumura14} 
There are two main points to be noted. 
(1) A staggered atomic displacement with $\bm{q}=(1,0,0)$ is induced in magnetic fields in phase II just below $T_{\text{N}}$, reflecting the CO of the $p$ band. 
(2) A parallel AFM ordering, where the moments are aligned parallel to the magnetic field, is induced in phase II, which can be understood by assuming the staggered ordering of the $\Gamma_7$ -- $\Gamma_8$ CF states. 
These situations are schematically illustrated in Fig.~1 of Ref.~\onlinecite{Matsumura14}. 
However, the details of the atomic displacements and the lattice distortion have remained unresolved. 
In the present paper, we report the x-ray diffraction studies on the atomic displacements and the lattice distortion in the ordered phases. 

This paper is organized as follows. 
In Sec.~II, we describe the details of the experimental methods. 
The first experiment is devoted to the determination of the atomic displacements of $\bm{q}=(1,0,0)$ in phase II and the second one to the detection of uniform lattice distortion by high precision measurement of the lattice parameter. 
The results and analysis of the former and the latter experiment is described in Sec.~III A and B, respectively. 
In Sec.~III A, a possible displacements of Ru and P atoms in phase II at 15 K and 6 Tesla are presented by assuming a $Pm\bar{3}$ space group, which is the same as in PrRu$_4$P$_{12}$ and PrFe$_4$P$_{12}$. 
In Sec.~III B, the results of the precise measurement of the lattice parameter is presented. From the splitting of the fundamental Bragg peaks, it is concluded that a rhombohedral distortion is induced along the $[1 1 1]$ axis, which coincides with the direction of the AFM moments. 
The variation in domain distribution as a function of magnetic field and temperature is also described, which is associated with the transition from parallel to perpendicular AFM structures. 
We discuss the results in Sec.~IV and a summary of the study is given in the final section.

\section{Experiment}
Single crystalline samples were grown by a tin-flux method. 
Two samples were prepared, one with the $(0\,1\,0)$ surface and the other with the $(1\,1\,1)$ surface.  
Both have approximately $1\times 1$ mm$^2$ flat surface area, which have been mirror polished. 
X-ray diffraction experiment has been performed at BL22XU in SPring-8. 
The samples were mounted in a 8 T vertical-field superconducting cryomagnet. 

The present x-ray diffraction experiment consists of two parts. One is the investigation of the Bragg peaks from atomic displacements induced by a magnetic field in phase II, which are forbidden in the paramagnetic phase. 
This experiment has been performed at a constant energy in a nonresonant condition by performing rocking scans ($\omega$-scans) and collecting the integrated intensities of the Bragg peaks. 
For $H\parallel [0\, 0\, 1]$, using the sample with the $(0\,1\,0)$ surface, the Bragg peaks in the $hk0$ scattering plane have been investigated. This is the same sample as we used in the previous work.\cite{Matsumura14,note1}
Using the sample with the $(1\,1\,1)$ surface, we investigated the Bragg peaks for $H\parallel [\bar{1}\, \bar{1}\, 2]$ and $H\parallel [1\, \bar{1}\, 0]$, where the scattering plane were spanned by $[1\,1\,1]$-$[1\,\bar{1}\,0]$ and by $[0\,0\,1]$-$[1\,1\,0]$, respectively. 

Another experiment is the high precision measurement of the lattice parameter. 
We use a high-resolution x-ray diffraction (HRXRD) method, consisting of a backscattering geometry of $2\theta \simeq 180^{\circ}$ and a high-resolution monochromator (HRM) system, which are aimed to improve the resolution of $\Delta d/d$ in the $\lambda = 2d \sin \theta$ relation. 
One of the usages of this method to a correlated electron system has been reported in Ref. \onlinecite{Inami14}, where the details of the method and the measurement system is explained. 
In general, the total resolution of $\Delta d/d$ is given by 
\begin{equation}
(\Delta d/d)^2 = (\cot \theta_{\text{\tiny B}}\, \Delta \theta)^2 + (\Delta \lambda/\lambda)^2 +(\Delta d_{\text{samp}}/d)^2\,,
\label{eq:1}
\end{equation}
where $\theta_{\text{\tiny B}}$ represents the Bragg angle and $\Delta d_{\text{samp}}$ the inhomogeneity in the inter-planar spacing $d$ caused by imperfections in the sample. 
The first term in  Eq.~(\ref{eq:1}) can be minimized by approaching the exact backscattering geometry of $\theta_{\text{\tiny B}} \rightarrow 90^{\circ}$. The second term is associated with the energy bandwidth of the incident beam. 
HRM is a second monochromator to reduce this term.  
We set a Si-660 channel-cut monochromator after the first double-crystal monochromator of Si-111, thereby further reducing the energy bandwidth from 1.3 eV to about 0.1 eV. 
Using HRM, at an x-ray energy of 10 keV, $\Delta \lambda / \lambda \simeq 1.0\times 10^{-5}$ is realized. 
However, if the third term, which depends on the quality of the sample, is larger than $\Delta \lambda / \lambda$, the total resolution of our experiment is determined by $\Delta d_{\text{samp}}/d$. 
In HRXRD experiment, we scan the x-ray energy by rotating HRM at a fixed Bragg angle $\theta_{\text{\tiny B}}$ close to $90^{\circ}$. This scan corresponds to a radial scan in the reciprocal space. 
The spatial distribution of the Bragg-peak intensity was measured by using an area detector PILATUS-100K, 
and the signal was integrated to obtain the total intensity.

\section{Results and Analysis}
\subsection{Field induced atomic displacements}
\begin{figure}[t]
\begin{center}
\includegraphics[width=8cm]{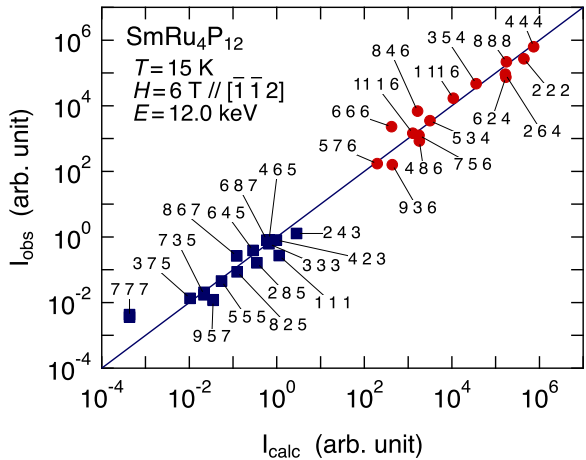}
\end{center}
\caption{(Color online) Comparison of the observed intensities with the calculated intensities for the Bragg reflections in phase II at 15 K and 6 T $\parallel [\bar{1}\, \bar{1}\, 2]$. The fundamental reflections and the forbidden reflections are shown by the circles and squares, respectively. }
\label{fig:Bragg112}
\end{figure}

\begin{figure}[t]
\begin{center}
\includegraphics[width=8cm]{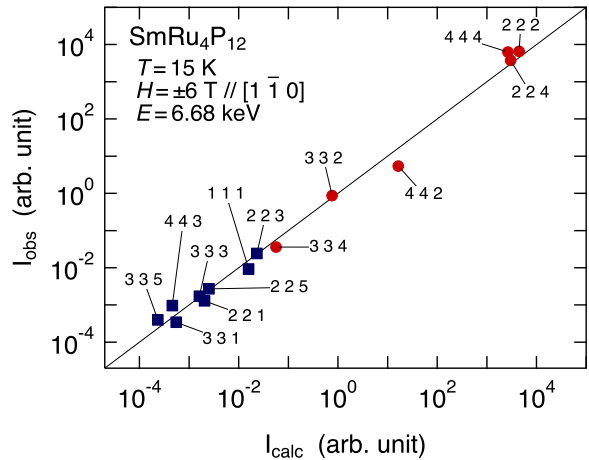}
\end{center}
\caption{(Color online) Comparison of the observed intensities with the calculated intensities for the Bragg reflections in phase II at 15 K and 6 T $\parallel [1\, \bar{1}\, 0]$. The data are averaged for reversed fields at $\pm 6$ T. The fundamental reflections and the forbidden reflections are shown by the circles and squares, respectively. }
\label{fig:Bragg110}
\end{figure}

\begin{figure}[t]
\begin{center}
\includegraphics[width=8cm]{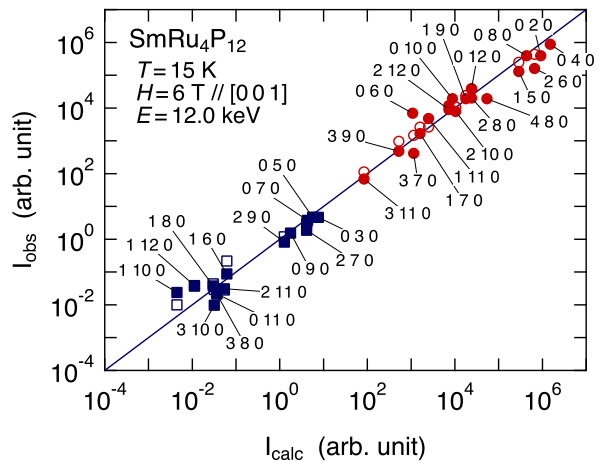}
\end{center}
\caption{(Color online) Comparison of the observed intensities with the calculated intensities for the Bragg reflections in phase II at 15 K and 6 T $\parallel [0\, 0\, 1]$. The fundamental reflections and the forbidden reflections are shown by the circles and squares, respectively. 
The closed and open marks represent the $hk0$ and $h\bar{k}0$ reflections, respectively, which have equal calculated intensities. The indices of the $h\bar{k}0$ reflections are omitted. }
\label{fig:Bragg001}
\end{figure}

We collected the intensities of the Bragg reflections at 15 K in a magnetic field of 6 T along $[0\, 0\, 1]$, $[\bar{1}\, \bar{1}\, 2]$, 
and $[1\, \bar{1}\, 0]$ directions and compared the integrated intensities with the calculated intensities assuming a model structure. 
The calculated intensity for a Bragg reflection at a scattering vector $\bm{Q} = \bm{k} - \bm{k}'$ is expressed by 
\begin{equation}
I_{\text{calc}}=\frac{S|F|^2\cos^2 2\theta}{\sin 2\theta} \,,
\label{eq:2}
\end{equation}
where $F=\sum_{j} (f_{0,j} + f'_{j} + if''_{j}) \exp (i\bm{Q} \cdot \bm{r}_{j})$ represents the structure factor, 
$2\theta$ the scattering angle, $\cos 2\theta$ the polarization factor for the $\pi$-$\pi'$ scattering process ($\bm{\varepsilon}, \bm{\varepsilon}' \perp \bm{k}\times\bm{k}'$), $\sin 2\theta$ the Lorentz factor, and $S$ is a constant scale factor.  
The scale factor was obtained by fitting the intensities of the fundamental Bragg reflections by assuming 
the well-determined atomic positions of $Im\bar{3}$ space group in the paramagnetic state; 
Sm ions are at the $2a$ site $(0,\, 0,\, 0)$, Ru at the $8c$ site $(\frac{1}{4},\, \frac{1}{4},\, \frac{1}{4})$, and 
P at the $24g$ site $(0,\, y,\, z)$ with $y=0.357$ and $z=0.1417$.\cite{Lee12}  
The results for the fundamental reflections are shown in Figs. \ref{fig:Bragg112}, \ref{fig:Bragg110}, and \ref{fig:Bragg001}. 
Note that there are no parameters to be refined here for these reflections. 
The scatter of the data around the line is caused by systematic errors in the present measurement.
The standard deviations as estimated by $\sigma=\{\sum_{i} (\log I_{\text{obs},i} - \log I_{\text{calc},i})^2 /N \}^{1/2}$ are 0.32, 0.28, and 0.28, for $H \parallel [0\, 0\, 1]$, $[\bar{1}\, \bar{1}\, 2]$, and $[1\, \bar{1}\, 0]$, respectively.  
The conventional $R_{\text{wp}}$$(S)$ factors are 0.32(1.3), 0.30(1.2), and 0.31(1.2), respectively.  
These values show the accuracy of the present experiment and are used to estimate the goodness of the fit for the forbidden reflections in the following. 
We note that the changes in intensity of the fundamental Bragg peaks between the ordered and paramagnetic phases are negligibly small in the scales of Figs. \ref{fig:Bragg112}, \ref{fig:Bragg110}, and \ref{fig:Bragg001}, and do not affect the $\sigma$ and $R_{\text{wp}}$ factors.

\subsubsection{A model of the field induced atomic displacements}
In this work, we try to fit the intensities of the forbidden reflections by assuming a space group $Pm\bar{3}$, which gives the same structure as the one reported in PrRu$_4$P$_{12}$.\cite{Lee04}  
Although this is a cubic space group and cannot be a solution for the structure in the AFM phase and in a magnetic field, where the cubic symmetry is broken, we adopt this model as a first step to interpret the experimental results. 
In the $Pm\bar{3}$ model, although the atomic positions of Sm do not change, the Sm sites at $(0,\, 0,\, 0)$ and $(\frac{1}{2},\, \frac{1}{2},\, \frac{1}{2})$ are called the $1a$ and the $1b$ site, respectively, indicating that the two Sm sites become inequivalent. 
Ru site changes from $(\frac{1}{4},\, \frac{1}{4},\, \frac{1}{4})$ to the $8i$ site at $(x,\, x,\, x)$. 
When we write $x=\frac{1}{4} + \delta$ ($\delta>0$), the Ru atoms shifts away from Sm-$1a$ and moves close to Sm-$1b$,  keeping the cubic symmetry. 
P site separates into the $12j$ site around Sm-$1a$ and the $12k$ site around Sm-$1b$. 
If we assume a symmetric displacement, the $12j$ site is expressed as $(0,\, y+\delta_u,\, z+\delta_v)$ and the $12k$ site as $(\frac{1}{2},\, y-\delta_u,\, z-\delta_v)$. 
Using Eq.~(\ref{eq:2}), we refined the three parameters to minimize the above defined standard deviation $\sigma$. This also gave the minimum $R_{\text{wp}}$ factor.
In PrRu$_4$P$_{12}$, the shift parameters are estimated as $\delta = 7 \times 10^{-4}$ for Ru 
and $(\delta_u, \delta_v) = (-3 \times 10^{-4}, 6 \times 10^{-4})$ for P.\cite{Lee04} 

In fitting the observed intensities of the forbidden reflections in magnetic fields, we first introduced the shift parameter $\delta$ for Ru. 
However, with the shifts of Ru only, the reflections of odd-only indices, such as 111, 331, or 735, vanishes.  
It is necessary to introduce shifts of both Ru and P. 
For $H\parallel [\bar{1}\, \bar{1}\, 2]$, by introducing $\delta_u = -0.5 \times 10^{-4}$ and $\delta_v = 1.3 \times 10^{-4}$ for P in addition to $\delta = 1.3 \times 10^{-4}$ for Ru, we could reasonably reproduce the observed intensities as shown in Fig.~\ref{fig:Bragg112}. 
%For $H\parallel [\bar{1}\, \bar{1}\, 2]$, by introducing $\delta_u = -0.50(5) \times 10^{-4}$ and $\delta_v = 1.26(5) \times 10^{-4}$ for P in addition to $\delta = 1.28(5) \times 10^{-4}$ for Ru, we could reasonably reproduce the observed intensities as shown in Fig.~\ref{fig:Bragg001}. 
%The number in the parenthesis represents the approximate error allowed for the last digit. 
The standard deviation of the fit, $\sigma=0.42$ $(R_{\text{wp}}=0.35, S=1.5)$, is as low as $\sigma=0.32$ $(R_{\text{wp}}=0.30, S=1.2)$ for the fundamental Bragg peaks. 

\begin{figure}[t]
\begin{center}
\includegraphics[width=8.5cm]{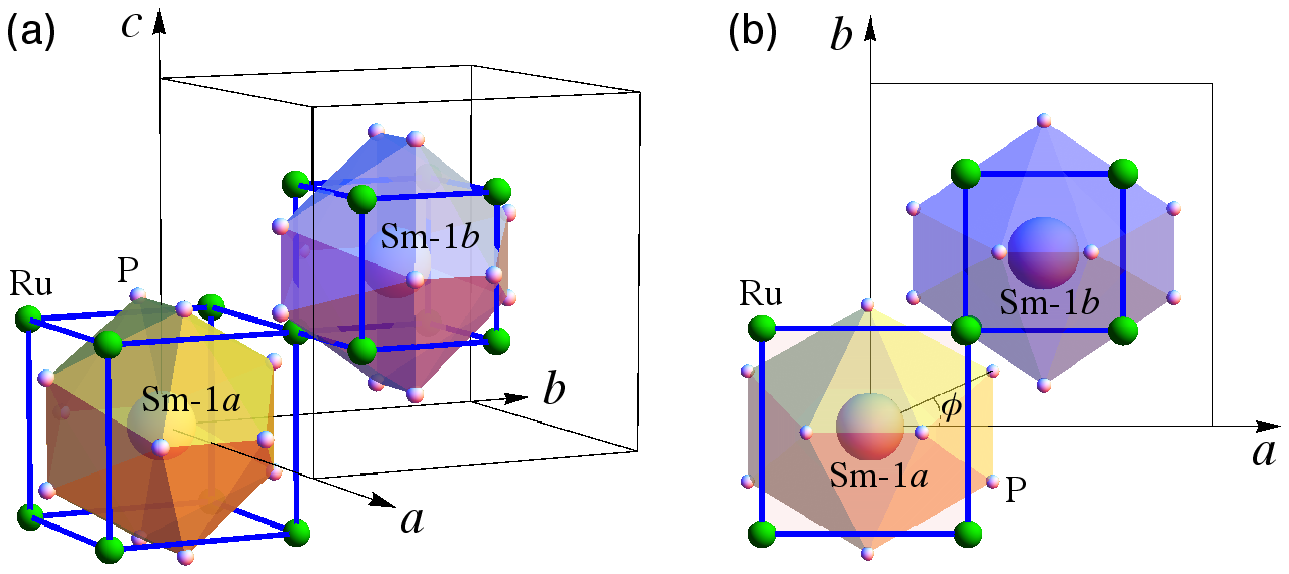}
\end{center}
\caption{(Color online) (a) Schematic view of the atomic displacements in phase II in magnetic fields, assuming a cubic $Pm\bar{3}$ space group. 
The displacements are emphasized than the real shifts. (b) Top view of (a). 
Around Sm-$1a$, the cube of Ru expands and the angle $\phi$ of P increases, whereas around Sm-$1b$ the atomic shifts are assumed opposite to those around Sm-$1a$. 
 }
\label{fig:structure}
\end{figure}

The obtained ratio of $\delta_u/\delta_v = -0.4$ is approximately equal to $-z/y=-0.397$ for the $24g$ site of P. 
This means that, when the P atom shifts, the angle $\phi$ between the $[1\, 0\, 0]$ axis and the position vector $(y,\, z,\, 0)$ changes, keeping the distance between Sm and P almost constant. 
This situation is shown in Fig.~\ref{fig:structure} in an exaggerated form of atomic shifts. 
Around Sm-$1a$, the cube of Ru expands, and the angle $\phi$ for P increases from 21.65$^{\circ}$ at $H=0$ to 21.67$^{\circ}$ at $H=6$ T for $H\parallel [\bar{1}\, \bar{1}\, 2]$.
Around Sm-$1b$, these shifts are opposite to those around Sm-$1a$. 

For $H\parallel [1\, \bar{1}\, 0]$, since the number of data points were small and we could not deduce reliable parameters by treating $\delta_u$ and $\delta_v$ independently, we fixed the relation of $\delta_u = -0.397 \delta_v$. 
Then, we obtained the best fit parameters of $\delta_u=-0.6 \times 10^{-4}$ and $\delta_v=1.5 \times 10^{-4}$. 
The standard deviation of the fit, $\sigma=0.20$ $(R_{\text{wp}}=0.23, S=1.1)$, is as small as $\sigma=0.28$ $(R_{\text{wp}}=0.31, S=1.2)$ for the fundamental Bragg peaks. 

For $H\parallel [0\, 0\, 1]$, it was possible to roughly reproduce the intensities only with the Ru-shift of $\delta=1.1\times 10^{-4}$ to $\sigma=0.44$. This is because all the observed forbidden reflections of $hk0$ are allowed by the shift of Ru. 
By introducing the shift of P, $\delta_u=-0.3 \times 10^{-4}$ and $\delta_v=0.7 \times 10^{-4}$ with the constraint of $\delta_u = -0.397 \delta_v$, $\sigma$ can be reduced to 0.32 $(R_{\text{wp}}=0.30, S=1.3)$, which is as small as $\sigma=0.28$ $(R_{\text{wp}}=0.32, S=1.3)$ for the fundamental Bragg peaks. 
The parameters used for the calculated intensities in Figs. \ref{fig:Bragg112}, \ref{fig:Bragg110}, and \ref{fig:Bragg001} are summarized in Table~\ref{tbl:1}.

Even if we could further reduce $\sigma$ and $R_{\text{wp}}$ by introducing more detailed model, it is beyond the accuracy of the present experiment and there is no meaning in deducing such detailed parameters. 
For $H\parallel [1\, \bar{1}\, 0]$, for example, if we consider a more realistic displacement of Ru, i.e., ($\delta+\delta'$, $\delta+\delta'$, $\delta$), the new parameter $\delta'$ slightly improves the fit. However, the improvement is much smaller than the accuracy of the present experiment.

\begin{table}
\caption{Atomic shift parameters of SmRu$_4$P$_{12}$ in phase II at 15 K and 6 T, assuming a cubic $Pm\bar{3}$ space group. 
%All the values have an allowed uncertainty of $\pm 0.05 \times 10^{-4}$. 
}
\label{tbl:1}
\renewcommand\arraystretch{1.5}
\begin{tabular}{ccccc}
\hline
 & & $H \parallel [\bar{1}\, \bar{1}\, 2]$ & $H \parallel [1\, \bar{1}\, 0]$ & $H \parallel [0 0 1]$ \\
 \hline
Ru & $\delta$ & $1.3 \times 10^{-4}$ & $1.3 \times 10^{-4}$ & $1.1 \times 10^{-4}$ \\
P & $\delta_u$ & $-0.5 \times 10^{-4}$ & $-0.6 \times 10^{-4}$ & $-0.3 \times 10^{-4}$ \\
 & $\delta_v$ & $1.3 \times 10^{-4}$ & $1.5 \times 10^{-4}$ & $ 0.7 \times 10^{-4}$ \\
%Ru & $\delta$ & $1.05 \times 10^{-4}$ & $1.28 \times 10^{-4}$ & $1.40 \times 10^{-4}$ \\
%P & $\delta_u$ & $-0.28 \times 10^{-4}$ & $-0.50 \times 10^{-4}$ & $-0.58 \times 10^{-4}$ \\
% & $\delta_v$ & $ 0.71 \times 10^{-4}$ & $1.26 \times 10^{-4}$ & $1.46 \times 10^{-4}$ \\
\hline
\end{tabular}
\end{table}

\begin{figure}[t]
\begin{center}
\includegraphics[width=8.5cm]{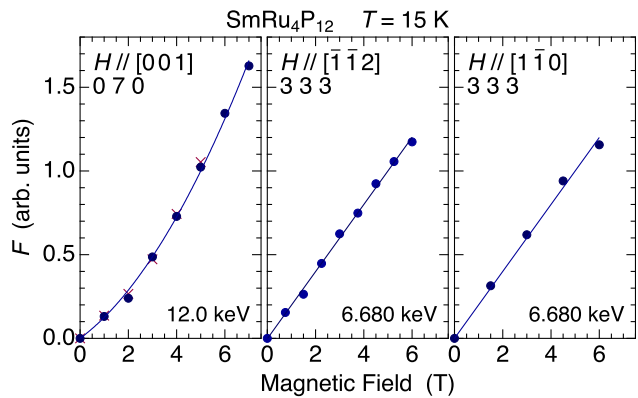}
\end{center}
\caption{(Color online) Magnetic field dependence of the structure factor of the forbidden reflections in phase II at 15 K. 
The data points were deduced by taking the square root of the integrated intensities of the rocking scans. 
The data are normalized to unity at 5 T. The crosses represent the data for the $0\,3\,0$ reflection reported in Ref.~\onlinecite{Matsumura14}. }
\label{fig:HdepFc}
\end{figure}

\subsubsection{Field dependence of the atomic displacement}
Figure \ref{fig:HdepFc} shows the magnetic field dependence of the structure factor of the forbidden reflections in phase II. 
Rocking ($\omega$) scans were performed and the square root of the integrated intensity is plotted as the structure factor. The data are normalized to unity at 5 T. 
For $H\parallel [\bar{1}\, \bar{1}\, 2]$ and $H\parallel [1\, \bar{1}\, 0]$, $F$ increases linearly with the field. 
For $H\parallel [0\, 0\, 1]$, on the other hand, $F$ shows a curve around 2 T, which well reproduces the previous data for $0\,3\,0$ shown by the crosses.\cite{Matsumura14} 
In Ref.~\onlinecite{Matsumura14}, we reported on this nonlinear increase above 2 T and connected it with the calculated field dependence of the order parameter $\phi_{Q}$ in Ref.~\onlinecite{Shiina14a}, where $\phi_{Q}$ shows a nonlinear increase before saturation. 
However, the present data does not exhibit a tendency to saturate in this field range up to 7 T. 
In view of the fact that the phase II region keeps expanding even at 30 T, this atomic displacement is expected to increase up to more higher fields above 30 T.\cite{Sekine03,Yoshizawa13} 
The nonlinear increase around 2 T, therefore, is a marginal behavior and should not be directly associated with the calculated nonlinear increase before saturation.

\subsection{High precision measurement of lattice parameter}
\subsubsection{$\bm{Q}=(8, 8, 8)$, $H\parallel [\bar{1}\, \bar{1}\, 2]$ }
\begin{figure}[t]
\begin{center}
\includegraphics[width=8.5cm]{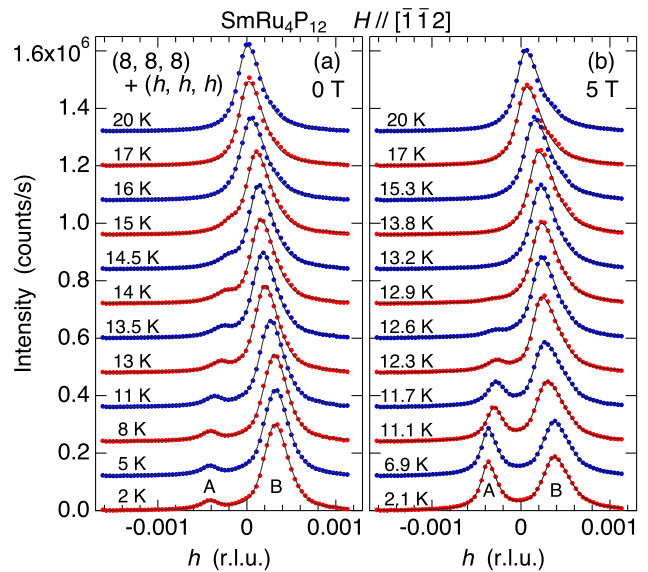}
\end{center}
\caption{(Color online) Radial scan profiles along $\bm{Q}=(8, 8, 8)+(h,h,h)$ at zero field (a) and at 5 T $\parallel [\bar{1}\, \bar{1}\, 2]$ (b). 
X-ray energy is 10.7055 keV at $h=0$. 
Solid lines are the fits using asymmetric pseudo-Voigt functions. }
\label{fig:prof888}
\end{figure}

\begin{figure}[t]
\begin{center}
\includegraphics[width=8.5cm]{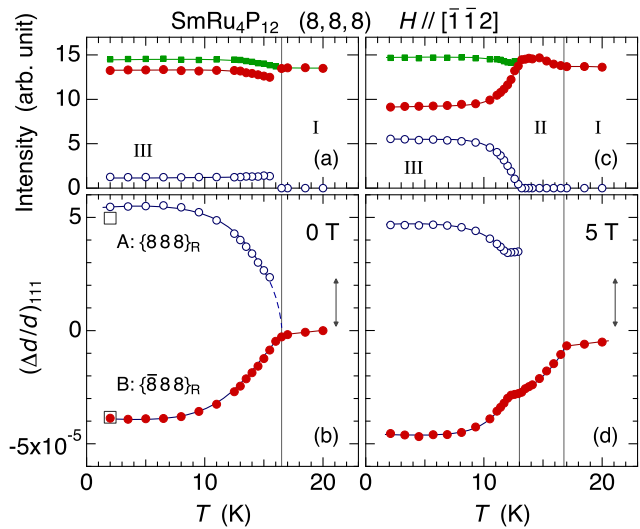}
\end{center}
\caption{(Color online) Temperature dependence of the integrated intensity and the relative variation in the planar spacing $(\Delta d/d)_{111}$ at zero field (a, b) and at 5 T (c, d), respectively, obtained from the fit of the data in Fig.~\ref{fig:prof888}. 
Open and closed circles correspond to peak A and B, respectively, in Fig.~\ref{fig:prof888}. 
Closed squares in (a) and (c) show the total intensity. The vertical arrows represent the HWHM of the peak profile at 20 K. 
Open squares in (b) are the calculated $(\Delta d/d)_{111}$ assuming a rhombohedral distortion (see text). 
}
\label{fig:dd888}
\end{figure}

Using the HRXRD system, energy scans at $\bm{Q}=(8, 8, 8)$ have been performed, corresponding to the radial scans in the reciprocal space along $(8+h,8+h,8+h)$. 
The results are shown in Fig.~\ref{fig:prof888}. 
At zero field, it is clearly demonstrated that the single peak in the paramagnetic phase splits into two peaks below $T_{\text{N}}$, 
indicating that the cubic symmetry is broken in the magnetic ordered phase. 
The peak profiles were fit with asymmetric pseudo-Voigt functions.  
The peak at the low-$Q$ and the high-$Q$ side is named A and B, respectively. 
Then, the integrated intensity and the relative variation in the planar spacing $(\Delta d/d)_{111}$ have been deduced as shown in Fig.~\ref{fig:dd888}. 

We can see that the intensity of the peak A is much weaker than that of peak B (only 8.6 \% of total). 
This is associated with the volume ratio of the structural domains contributing to peak A and B. 
The total intensity slightly increases below $T_{\text{N}}$ probably because the extinction effect is reduced by the lattice distortion. 
Although the peak splitting seems to vanish continuously at $T_{\text{N}}$, 
it is difficult to identify the splitting in the temperature region just below $T_{\text{N}}$ because the peak A is hidden in the tail of the main peak. 
The half-width-at-half-maximum (HWHM) of the peak, as transformed to HWHM of $(\Delta d/d)_{111}$, is $2.1\times 10^{-5}$ at 20 K as indicated by the vertical arrow in Fig.~\ref{fig:dd888}(b). 
If the splitting is less than this HWHM, it is difficult to find the correct peak position by the fitting. 
We cannot mention from the present data whether or not $(\Delta d/d)_{111}$ of peak A continuously decreases to zero at $T_{\text{N}}$. 
A speculated line of $(\Delta d/d)_{111}$ for peak A at 0 T is represented by the dashed line in Fig.~\ref{fig:dd888}(b). 

In a magnetic field of 5 T along $[\bar{1}\, \bar{1}\, 2]$, as shown in Fig.~\ref{fig:prof888}(b), the intensity of peak A at 2 K increases significantly whereas that of peak B decreases in comparison with those at 0 T. 
This shows that the structural domain of peak A is more favored by applying a magnetic field in phase III. 
The temperature dependences of the intensity and $(\Delta d/d)_{111}$ are shown in Fig.~\ref{fig:dd888}(c) and \ref{fig:dd888}(d), respectively. 
It is very surprising that the peak remains in a single peak in phase II below $T_{\text{N}}$=16.8 K, 
where the peak shows a shift to the high-$Q$ side with decreasing temperature. 
Furthermore, it is also interesting that the peak A appears in phase III at a well separated position from peak B and the intensity increases from zero. 
This can be understood by considering that the structural domain of peak A do not exist in phase II in magnetic fields and starts to develop on entering phase III. 

\subsubsection{$\bm{Q}=(8, 8, 8)$, $H\parallel [1\, \bar{1}\, 0]$ }
Almost the same magnetic-field effect is observed also for $H\parallel [1\, \bar{1}\, 0]$ (the data are not shown). 
The intensity of peak A at 2 K in phase III increases by applying a magnetic field in the same manner as for $H\parallel [\bar{1}\, \bar{1}\, 2]$. 
This shows that the structural domain of peak A is favored in magnetic fields in phase III. 
In phase II, the (8, 8, 8) peak remains a single peak and shifts to the high-$Q$ side with decreasing temperature. 
On entering phase III, the peak A appears at a well separated position from peak B and the intensity increases from zero. 
All these behaviors can be understood by considering that the structural domain of peak A do not exist in phase II, and it starts to develop on entering phase III. 

\subsubsection{$\bm{Q}=(10, 10, 0)$, $H\parallel [0\, 0\, 1]$}
\begin{figure}[t]
\begin{center}
\includegraphics[width=8.5cm]{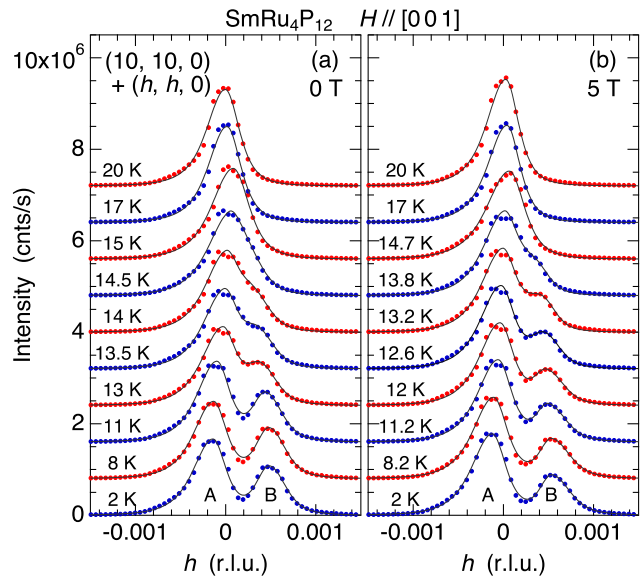}
\end{center}
\caption{(Color online) Radial scan profiles along $\bm{Q}=(10, 10, 0)+(h, h, 0)$ at zero field (a) and at 5 T $\parallel [0 0 1]$ (b). 
X-ray energy is 10.9263 keV at $h=0$. 
Solid lines are the fits using asymmetric pseudo-Voigt functions. }
\label{fig:Prof10100}
\end{figure}

\begin{figure}[t]
\begin{center}
\includegraphics[width=8.5cm]{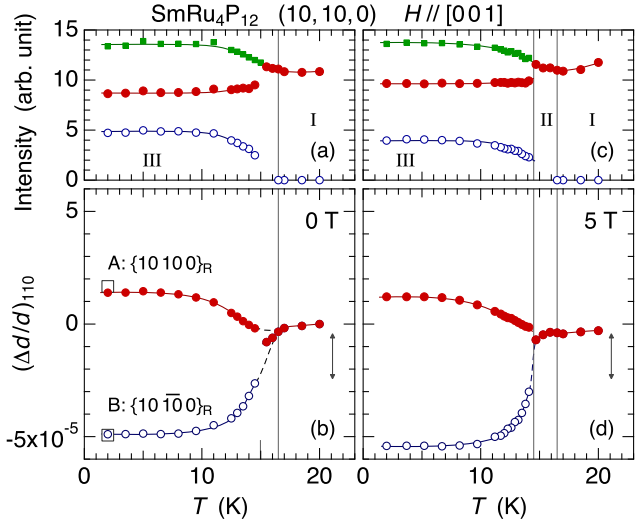}
\end{center}
\caption{(Color online) Temperature dependence of the integrated intensity and the relative variation in the planar spacing $(\Delta d/d)_{110}$ at zero field (a, b) and at 5 T (c, d), respectively, obtained from the fit of the data in Fig.~\ref{fig:Prof10100}. 
Open and closed circles correspond to peak A and B, respectively, in Fig.~\ref{fig:Prof10100}. 
Closed squares in (a) and (c) show the total intensity. Vertical arrows represent the HWHM of the peak profile at 20 K.
Open squares in (b) are the calculated $(\Delta d/d)_{110}$ assuming a rhombohedral distortion (see text). }
\label{fig:dd110}
\end{figure}

Figure~\ref{fig:Prof10100} shows the radial profiles along $\bm{Q}=(10, 10, 0)$. 
Also in this geometry, it is clearly demonstrated that the single peak in the paramagnetic phase splits into two peaks in phase III. 
The temperature dependences of the integrated intensities and the relative variation in the planar spacing $(\Delta d/d)_{110}$ are shown in Fig.~\ref{fig:dd110}. 
The intensities of the two peaks are in the same order of magnitude, which is in contrast to the case for $(8, 8, 8)$.

At zero field, strangely, the peak does not seem to split below $T_{\text{N}}$=16.5 K, but seems to split below 15 K. 
Therefore, the data points between 15 K and 16.5 K in Fig.~\ref{fig:dd110}(b) was obtained by assuming a single peak. 
However, as shown by the vertical arrow in Fig.~\ref{fig:dd110}(b), the HWHM of the peak profile is larger than the expected peak splitting in this temperature range between 15 K and 16.5 K. 
It is difficult to find the peak splitting even if it exists. 
Since the peak splitting is actually observed in the scan at $(8,\, 8,\, 8)$ below $T_{\text{N}}$, it is reasonable to consider that the peak is also split in this scan at $(10,\, 10,\, 0)$. 
The $T$-dependences of $(\Delta d/d)_{110}$ just below $T_{\text{N}}$, which we speculate, are shown by the dashed lines.  

The intensities of peak A and B do not change much by applying a magnetic field of 5 T along $[0\, 0\, 1]$. 
This is in contrast to the case for $(8,\, 8,\, 8)$ with $H\parallel [\bar{1}\, \bar{1}\, 2]$ and $H\parallel [1\, \bar{1}\, 0]$, where the intensity of peak B increased significantly. 
We consider that this reflects the response of the domain population to the magnetic field, which will be discussed later in association with the magnetic structure. 
The peak splitting in phase II at 5 T is also difficult to mention in this geometry. The data in Figs.~\ref{fig:dd110}(b) and \ref{fig:dd110}(d) look similar. If we look carefully, however, an extrapolation of $(\Delta d/d)_{110}$ for peak B to the phase boundary, as shown by the dashed line in Fig.~\ref{fig:dd110}(d), suggests that the splitting vanishes at a lower temperature than $T_{\text{N}}$, i.e., the phase II exists below $T_{\text{N}}$.

\subsubsection{$\bm{Q}=(0, 14, 0)$, $H\parallel [0\, 0\, 1]$}
\begin{figure}[t]
\begin{center}
\includegraphics[width=8.5cm]{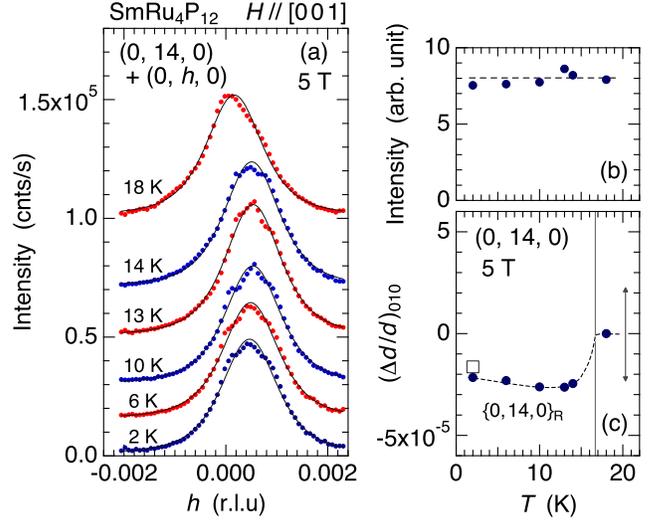}
\end{center}
\caption{(Color online) 
(a) Radial scan profiles along (14, 0, 0)$+(h, 0, 0)$ at 5 T  $\parallel [0\, 0\, 1]$. 
X-ray energy is 10.8189 keV at $h=0$. 
Solid lines are the fits using an asymmetric squared Lorentzian function. 
(b) and (c) show the temperature dependence of the integrated intensity and $(\Delta d/d)_{010}$ obtained from the fit, respectively. 
Open square in (c) is the calculated $(\Delta d/d)_{010}$ assuming a rhombohedral distortion (see text). }
\label{fig:dd010}
\end{figure}

Figure~\ref{fig:dd010}(a) shows the peak profiles at (0, 14, 0) in a magnetic field of 5 T along $[0\, 0\,  1]$. 
The temperature dependences of the integrated intensity and the relative variation of $(\Delta d/d)_{010}$ are also shown in Figs.~\ref{fig:dd010}(b) and \ref{fig:dd010}(c), respectively. 
Although the HWHM of the peak profile is larger than those of other reflections, we can see that the profile remains a single peak down to the lowest temperature of 2 K because the peak width does not change. 
This shows that all the domains in phase III in the lowered crystal symmetry has the same $(0\,1\,0)$ inter-planar spacing. 
The magnetic field does not change this situation. 
Although the number of data points is small, we can clearly see that $(\Delta d/d)_{010}$ decreases below $T_{\text{N}}$. 

\subsubsection{Rhombohedral distortion}
From the result that the $(8, 8, 8)$ and $(10, 10, 0)$ reflections split into two peaks, whereas the $(14, 0, 0)$ reflection remains a single peak, we can conclude that the crystal symmetry in phase III is rhombohedral. 
The crystal expands or contracts along one of the four $\langle 1\, 1\, 1 \rangle$ directions. 
If the crystal expands (contracts) along the $[1\, 1\, 1]$ direction, it contracts (expands) in the $[\bar{1}\,1\,1]$, $[1\,\bar{1}\,1]$, and $[1\, 1\, \bar{1}]$ directions. 
The cubic $\{1\, 1\, 1\}$ planes split into rhombohedral $\{1\, 1\, 1\}_{\text{R}}$ and $\{\bar{1}\, 1\, 1\}_{\text{R}}$ planes. 
The indices without subscripts refer to the original cubic lattice and those with the subscript R refer to the rhombohedral lattice. 
The splitting results in producing four rhombohedral domains and two different $d$-spacings for cubic $(h, h, h)$ and $(h, h, 0)$ reflections. 
We need to determine which of the peak A and B does the rhombohedral $(8, 8, 8)_{\text{R}}$ reflection belong to.

Cubic to rhombohedral deformation can be expressed by two parameters $\varepsilon_{\text{\tiny B}}$ and $\varepsilon_{xy}$.\cite{Kubo04} 
The former expresses an isotropic volume expansion, 
\begin{equation}
\varepsilon_{\text{\tiny B}} = \varepsilon_{xx} +  \varepsilon_{yy} +  \varepsilon_{zz} \,, 
\end{equation}
and the latter expresses a shear strain, 
\begin{equation}
\varepsilon_{xy} = \varepsilon_{yz} = \varepsilon_{zx}\,,
\end{equation}
where the strain tensor $\varepsilon_{\mu\nu}$ ($\mu, \nu$ = $x, y, z$) is defined as \cite{Kittel}
\begin{equation}
\varepsilon_{\mu\nu} = \frac{1}{2}\left( \frac{\partial u_{\mu}}{\partial r_{\nu}} 
+ \frac{\partial u_{\nu}}{\partial r_{\mu}} \right)\, . 
\end{equation}
The relative length change $\Delta l/l$ along the direction defined by the direction cosine $(\alpha,\beta,\gamma)$ is calculated as 
\begin{align}
\frac{\Delta l}{l} &= \varepsilon_{xx} \alpha^2 + \varepsilon_{yy} \beta^2 + \varepsilon_{zz} \gamma^2 \nonumber \\
& \;\;\;\; + 2(\varepsilon_{yz}\beta\gamma + \varepsilon_{zx}\gamma\alpha + \varepsilon_{xy}\alpha\beta)\,.
\end{align}
Then, using $\varepsilon_{\text{\tiny B}}$ and $\varepsilon_{xy}$, the cubic to rhombohedral lattice deformation can be expressed as the following: 
\begin{align}
(\Delta d/d)_{111} &= \frac{1}{3}\varepsilon_{\text{\tiny B}} + 2\varepsilon_{xy} \,,\nonumber \\
(\Delta d/d)_{\bar{1}11} &= \frac{1}{3}\varepsilon_{\text{\tiny B}} - \frac{2}{3}\varepsilon_{xy} \,,\nonumber \\
(\Delta d/d)_{110} &= \frac{1}{3}\varepsilon_{\text{\tiny B}} +  \varepsilon_{xy} \,,\label{eq:6} \\
(\Delta d/d)_{1\bar{1}0} &= \frac{1}{3}\varepsilon_{\text{\tiny B}} -  \varepsilon_{xy} \,,\nonumber \\
(\Delta d/d)_{010} &= \frac{1}{3}\varepsilon_{\text{\tiny B}} \,.\nonumber 
\end{align}
By putting the experimental values of $\Delta d/d$ at 2 K in Figs.~\ref{fig:dd888} and \ref{fig:dd110}, we can estimate $\varepsilon_{\text{\tiny B}}$ and $\varepsilon_{xy}$ at the lowest temperature. 
If we assume the rhombohedral $(8, 8, 8)_{\text{R}}$ reflection belong to peak B in Fig.~\ref{fig:prof888}, i.e., the crystal expands along one of the $[\bar{1}\,1\,1]$, $[1\,\bar{1}\,1]$, and $[\bar{1}\,\bar{1}\,1]$ directions, we do not have consistent parameters of $\varepsilon_{\text{\tiny B}}$ and $\varepsilon_{xy}$. 
Only by assuming that the peak A in Fig.~\ref{fig:prof888} corresponds to the rhombohedral $(8, 8, 8)_{\text{R}}$ reflection, 
i.e., by assuming that the crystal expands along [111], and the peak B corresponds to the superposition of $(\bar{8}, 8, 8)_{\text{R}}$, $(8, \bar{8}, 8)_{\text{R}}$, and $(8, 8, \bar{8})_{\text{R}}$ reflections, the data can be explained consistently. 
By assuming 
\begin{equation}
\varepsilon_{\text{\tiny B}} = -4.7 \times 10^{-5}\;\; \text{and}\;\; \varepsilon_{xy} = 3.3 \times 10^{-5}\,, \nonumber
\end{equation}  
the $\Delta d /d$ values are calculated as 
$(\Delta d/d)_{111}$=$5.0 \times 10^{-5}$, 
$(\Delta d/d)_{11\bar{1}}$=$-3.8 \times 10^{-5}$, 
$(\Delta d/d)_{110}$=$1.7 \times 10^{-5} $,
$(\Delta d/d)_{1\bar{1}0}$=$-4.9 \times 10^{-5}$, and 
$(\Delta d/d)_{010}$=$-1.6 \times 10^{-5}$, 
which agree well with the $\Delta d/d$ values at the lowest temperature of 2 K as shown by the open squares in Figs.~\ref{fig:dd888}, \ref{fig:dd110}, and \ref{fig:dd010}. 
These results confirm that the crystal symmetry in phase III is rhombohedral and the unit cell expands along the $[1\, 1\, 1]$ direction. 
This is consistent with the result of NQR analysis.\cite{Masaki07}  
The above strain parameters of $\varepsilon_{\text{\tiny B}}$ and $\varepsilon_{xy}$ correspond to the rhombohedral angle of $89.9962^{\circ}$ and the relative change in the lattice parameter $\Delta a/a = -1.6 \times 10^{-5}$. 
The rhombohedral deformation is caused through magneto-elastic coupling and is associated with the direction of the antiferromagnetic moments, which is also along the $[1\, 1\, 1]$ direction.\cite{Lee12,Aoki07} 

The intensity of peak A for the rhombohedral $(8, 8, 8)_{\text{R}}$ reflection at zero field in Fig.~\ref{fig:prof888} should ideally be 1/3 to that of peak B consisting of three equivalent $(\bar{8}, 8, 8)_{\text{R}}$ reflections. 
However, the ratio is much weaker than the ideal value. 
This can be ascribed to an anisotropic stress on the $(1\,1\,1)$ sample surface, which probably works to expand the surface along the directions perpendicular to the $[1\,1\,1]$ axis and suppresses the development of peak A. 
The intensities of the two peaks in the $(10, 10, 0)$ reflection of Fig.~\ref{fig:Prof10100}, on the other hand, is more consistently distributed. This can also be understood by considering that the anisotropic stress on the $(0\,1\,0)$ sample surface does not lead to the imbalance of domain population. 

\subsubsection{Field response of rhombohedral domains}
\begin{figure}[t]
\begin{center}
\includegraphics[width=8.5cm]{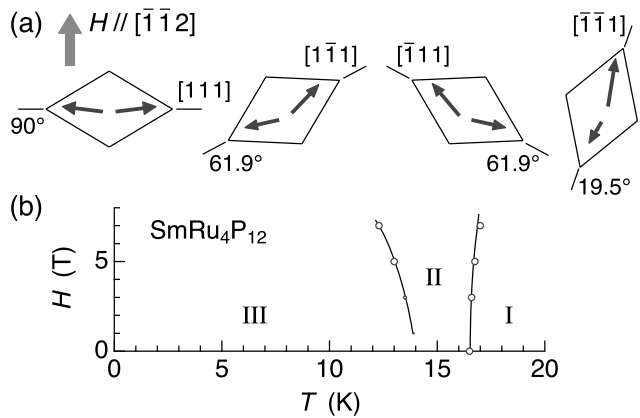}
\end{center}
\caption{
(a) Schematic of the rhombohedral domains and the AFM moments in a magnetic field along $[\bar{1}\, \bar{1}\, 2]$. 
The angles between the magnetic field and the rhombohedral principal axis are also shown. 
(b) Magnetic phase diagram of SmRu$_4$P$_{12}$ for $H\parallel [\bar{1}\, \bar{1}\, 2]$ obtained in the present work. 
The $[\bar{1}\, \bar{1}\, 1]$ domain is favored in phase II and the $[1\, 1\, 1]$ domain in phase III, respectively.  
}
\label{fig:domain}
\end{figure}

If we apply a magnetic field in phase III along the $[0\, 0\, 1]$ direction, all the AFM domains ordered along the four equivalent $\langle 1\, 1\, 1 \rangle$ directions have equal magnetic energies. 
Therefore, the domain population is expected not to change much by the field. 
Since the principal axis of the rhombohedral distortion coincides with the direction of the AFM moment, the intensities of peak A and B in Fig.~\ref{fig:dd110} also do not change much by a magnetic field applied along $[0\, 0\, 1]$. 
By contrast, for $H \parallel [\bar{1}\, \bar{1}\, 2]$, the AFM moments in the $[1\, 1\, 1]$ domain are perpendicular to $H$, those in the $[\bar{1}\, 1\, 1]$ and $[1\, \bar{1}\, 1]$ domains have an angle of 62$^{\circ}$ with $H$, and those in the $[\bar{1}\, \bar{1}\, 1]$ domain have an angle of 20$^{\circ}$. 
This situation is shown in Fig.~\ref{fig:domain}. 
Therefore, in normal cases, the slightly canted AFM moments in the $[1\, 1\, 1]$ domain (perpendicular AFM) has the lowest magnetic energy among the four domains. 
This is the reason that the intensity of the peak A in Fig.~\ref{fig:prof888}, corresponding to the rhombohedral $(8, 8, 8)_{\text{R}}$ reflection, increases with increasing the field. 
The intensity at 2 K, although it is only 8.6\% of total at $H=0$, increases to 36\% at 5 T, and at 7 T, it overcomes the intensity of peak B and reaches to 51 \% of total. 
In the same manner, for $H \parallel [1\, \bar{1}\, 0]$, the $[1\, 1\, 1]$ and $[1\, \bar{1}\, 1]$ domains are favored in phase III because the AFM moments in these domains are perpendicular to $H$, whereas those in the $[\bar{1}\, 1\, 1]$ and $[1\, \bar{1}\, 1]$ domains have an angle of 35.3$^{\circ}$ with $H$.

It is more remarkable that the $(8, 8, 8)$ reflection is single peaked in phase II for $H \parallel [\bar{1}\, \bar{1}\, 2]$ and also for $H \parallel [1\, \bar{1}\, 0]$. 
As shown in Fig.~\ref{fig:dd888}, the peak A, corresponding to the $[1\, 1\, 1]$ domain preferred in phase III, does not exist in phase II. 
The single peak B shows contraction in the $[1\, 1\, 1]$ direction. 
This indicates that a rhombohedral domain is selected so that the elongated principal axis is oriented to the direction of the magnetic field. 
That is, only the $[\bar{1}\, \bar{1}\, 1]$ domain in Fig.~\ref{fig:domain} is selected for $H \parallel [\bar{1}\, \bar{1}\, 2]$. 
In the same manner, the $[1\, \bar{1}\, 1]$ and $[\bar{1}\, 1\, 1]$ domains are selected for $H \parallel [1\, \bar{1}\, 0]$. 
This is consistent with the previous report of resonant x-ray diffraction that the AFM moments in phase II are oriented parallel to the magnetic field. 
This is the anomalous AFM state (parallel AFM) peculiar to the field-induced CO phase in this compound. 

Note that the rhombohedral symmetry in phase II has not been confirmed in a strict sense in the present experiment because only a single peak has been detected. 
However, from almost the same temperature dependence of $(\Delta d/d)_{111}$ for peak B at 5 T as that at 0 T, as shown in Fig.~\ref{fig:dd888}, we may infer that the crystal symmetry in phase II is rhombohedral at least for $H \parallel [\bar{1}\, \bar{1}\, 2]$ and for $H \parallel [1\, \bar{1}\, 0]$.
%For $H \parallel [0\, 0\, 1]$, since $(\Delta d/d)_{110}$ for the single peak in phase II at 5 T shows little temperature dependence as shown in Fig.~\ref{fig:dd110}, 

Another point to be noted is that the peak of the $[1\, 1\, 1]$ domain appears at a clearly separated position on entering phase III whereas the intensity continuously increases from zero. 
This probably shows that the AFM moments flips from parallel to perpendicular configuration with respect to the field on entering the low-$T$ phase III. 
Since the $\Delta d/d$ value reflects the magnitude of the ordered moment, the $(8, 8, 8)$ peak appears at a separated position. On the other hand, the intensity is proportional to the volume fraction of the flipped region, resulting in a continuous increase from zero intensity. 
Although this is a first order transition, no hysteresis was observed in the cooling and heating processes.

\section{Discussion}
In Sec.~III A, we deduced the atomic shift parameters in phase II by assuming a cubic $Pm\bar{3}$ space group. 
In Sec.~III B, on the other hand, we showed that the crystal symmetry is rhombohedral in phase III and that it is also expected to be the case in phase II. 
The correct space group should therefore be $R\bar{3}$ instead of $Pm\bar{3}$. 
However, the rhombohedral distortion is so small that it is not resolved in the data in Sec.~III A. 
The observed intensity involve reflections from all the rhombohedral domains. 
Therefore, there is little meaning to adopt the $R\bar{3}$ space group to analyze the data in Sec.~III A. 
In the $R\bar{3}$ space group, the Ru sites split into the $2c$ site at $(x,x,x)$ and the general $6f$ site at $(x,y,z)$. 
All the P atoms also belong to the $6f$ site. The determination of these shift parameters of the $6f$ site, however, is too detailed and beyond the accuracy of the present experiment as described in Sec.~III A. 
The $Pm\bar{3}$ model is sufficient to interpret the field-induced atomic displacements in phase II. 

One of the fundamental problems in phase II is that which of Sm-$1a$ and Sm-$1b$ the conduction $p$ electrons gather around. 
Unfortunately, the present experiment provides no direct information on the charge density of the $p$ electrons. 
A simplistic consideration may be that the lattice expands around Sm where the charge density is large, i.e., the charge density increases around Sm-$1a$ in the model of Fig.~\ref{fig:structure}. 
This problem is of fundamental importance because the charge density is associated with the ground state nature of the Sm $4f$ state, i.e., $\Gamma_7$-like or $\Gamma_8$-like. 
Theoretically, the charge density and the magnetic moment at Sm sites with the $\Gamma_7$-like ground state will be larger than those at Sm sites with the $\Gamma_8$-like ground state, which is illustrated schematically in Fig. 1 of Refs.~\onlinecite{Shiina13} and \onlinecite{Matsumura14}. 
This alternate arrangement of the charge density and the CF states results in the total energy gain and the anomalous AFM structure of long and short moments oriented parallel to the magnetic field in phase II. 
The relationship between the magnitude of the magnetic moment and the atomic displacements around Sm, or the charge density, should be clarified experimentally in future. 

The atomic shift parameters summarized in Table~\ref{tbl:1} show that the field-induced atomic displacements of Ru and P in phase II do not depend much on the field direction. 
This result suggests that the charge density and the CF states also do not change with the field direction. 
This fact can be associated with the totally symmetric order parameter of phase II. 
This point will be further studied by measuring the field-direction dependence of the parallel AFM order by resonant x-ray diffraction. 

The boundary between phase II and III become less prominent with decreasing magnetic field and seems to disappear at zero field.
This is associated with the fact that the atomic displacement and the parallel AFM, which is  characteristic in phase II, is induced almost linearly with the applied field as shown in Fig.~\ref{fig:HdepFc}. 
The difference between the parallel AFM in phase II and the perpendicular AFM in phase III is significant at high fields, but it is small at low fields and vanishes at zero field. 
%At zero field, the AFM ordered state consists of four magnetic domains oriented along the $\langle 1\,1\,1\rangle$ axes with rhombohedral distortions. This is a normal AFM order. 
Therefore, it is suggested that the II-III boundary disappears and the phase II does not exist at zero field, 
although the II-III boundary seems to approach $T^*\sim 14$ K at zero field in Fig.~\ref{fig:domain}(b). 
It could also be stated that the CO and the parallel AFM state of phase II appears only in a magnetic field, no matter how small it is. At zero field, only the normal AFM order exists. 
This is also associated with the upturn anomaly in the magnetic susceptibility on entering phase II from phase I, which is expected to appear down to small magnetic fields below 1 T and disappears at zero field.\cite{Sekine03} 
Consistency with the theoretical phase diagram should be studied experimentally at low fields below 1 T.  

Although the possibility of multipolar moments to participate in the ordering phenomenon in SmRu$_4$P$_{12}$ is not completely discarded, we consider that the present picture of field-induced CO is more consistent with the experimental results. 
As studied in Ref.~\onlinecite{Kiss09}, a similar phase diagram can be reproduced by considering a mixed order of $\Gamma_{5u}$ octupole and $\Gamma_{4g}$ hexadecapole, where the boundary at $T^*$ is interpreted as a crossover with a Schottky anomaly. 
However, the signal of resonant scattering reported in Ref.~\onlinecite{Matsumura14} is consistently explained by magnetic dipole, and not by $\Gamma_{5u}$ octupole or by $\Gamma_{4g}$ hexadecapole. 
Magnetic dipole moment of 0.3 $\mu_{\text{B}}$ observed in the experiment\cite{Aoki07,Lee12} is more consistently interpreted as the magnetic dipole of the $\Gamma_7$ CF-state (0.24 $\mu_{\text{B}}$), but is difficult to explain by the $\Gamma_{5u}$ octupole ordering model, which gives extremely small dipole moment.\cite{Kiss09} 
The field-induced atomic displacement clarified in the present work ($Pm\bar{3}$ model) is more consistent with the $\Gamma_{1g}$ order parameter, i.e., the CO or the hexadecapole order of total symmetry, than the $\Gamma_{4g}$ hexadecapole. 
As clearly shown in Fig.~\ref{fig:dd888}, the boundary at $T^*$ is not a crossover, but a phase transition between the CO with parallel AFM (phase II) and the perpendicular AFM (phase III).

\section{Summary}
We have performed nonresonant x-ray diffraction experiments to clarify the staggered atomic displacements of Ru and P with $\bm{q}=(1,0,0)$ in the field-induced charge ordered state in SmRu$_4$P$_{12}$ realized in the temperature region of $T^* < T < T_{\text{N}}$. 
The displacement parameters of Ru and P for the field directions of $[\bar{1}\, \bar{1}\, 2]$, $[1\, \bar{1}\, 0]$, and $[0\, 0\, 1]$ have been deduced by assuming a cubic space group $Pm\bar{3}$, which is sufficient as a first step analysis.  
The cube of Ru atoms around Sm expands and shrinks alternately and the P atoms shift in accordance with the Ru shifts. 
These shifts must be associated with the different charge densities of $p$ electrons around the Sm atoms. 

From the high precision measurement of the lattice parameter, we confirmed that a rhombohedral distortion takes place below $T_{\text{N}}$. At zero field without charge order, the lattice is elongated along the principal axis of $[1\, 1\, 1]$, along which the AFM moments are aligned.  
In magnetic fields, in the low temperature phase below $T^*$, the rhombohedral principal $[1\, 1\, 1]$ axis prefers to be perpendicular to the applied field. This is a normal AFM state. 
On the other hand, in the field-induced charge ordered phase above $T^*$, the rhombohedral principal $[1\, 1\, 1]$ axis prefers to be parallel to the applied field. 
In the parallel AFM state, the long and short magnetic moments are ordered, which is consistent with the theoretical prediction of the alternating arrangement of the $\Gamma_7$ -- $\Gamma_8$ CF states.

The authors acknowledge valuable discussions with R. Shiina. This work was supported by JSPS KAKENHI Grant Numbers 2430087 and 15K05175. The synchrotron radiation experiment was performed under Proposal numbers 2014A3711, 2014B3711, and 2015A3711 at BL22XU of SPring-8.

%\bibliography{dfcoul_prl}

\begin{thebibliography}{99}
\bibitem{note1} The notation of $\bm{Q}=(3,0,0)$ and $H\parallel [001]$ in Ref.~\onlinecite{Matsumura14} should be changed to $\bm{Q}=(0,3,0)$ and $H\parallel [001]$ (or $\bm{Q}=(3,0,0)$ and $H\parallel [010]$) 
because they are not equivalent in a strict sense in the crystal structure of the filled skutterudite compounds (space group $Im\bar{3}$), where the $[100]$ is not a four fold axis. However, this does not affect the analysis and the discussion of Ref.~\onlinecite{Matsumura14}.
\end{thebibliography}

\end{document}